\documentclass[useAMS,usenatbib]{mn2e}
\usepackage{multirow}
\usepackage{epsfig}
\usepackage{subfigure}
\usepackage{url}

\makeatletter
\let\jnl@style=\rm
\def\ref@jnl#1{{\jnl@style#1}}

\def\aj{\ref@jnl{AJ}}                   
\def\araa{\ref@jnl{ARA\&A}}             
\def\apj{\ref@jnl{ApJ}}                 
\def\apjl{\ref@jnl{ApJ}}                
\def\apjs{\ref@jnl{ApJS}}               
\def\ao{\ref@jnl{Appl.~Opt.}}           
\def\apss{\ref@jnl{Ap\&SS}}             
\def\aap{\ref@jnl{A\&A}}                
\def\aapr{\ref@jnl{A\&A~Rev.}}          
\def\aaps{\ref@jnl{A\&AS}}              
\def\azh{\ref@jnl{AZh}}                 
\def\baas{\ref@jnl{BAAS}}               
\def\jrasc{\ref@jnl{JRASC}}             
\def\memras{\ref@jnl{MmRAS}}            
\def\mnras{\ref@jnl{MNRAS}}             
\def\pra{\ref@jnl{Phys.~Rev.~A}}        
\def\prb{\ref@jnl{Phys.~Rev.~B}}        
\def\prc{\ref@jnl{Phys.~Rev.~C}}        
\def\prd{\ref@jnl{Phys.~Rev.~D}}        
\def\pre{\ref@jnl{Phys.~Rev.~E}}        
\def\prl{\ref@jnl{Phys.~Rev.~Lett.}}    
\def\pasp{\ref@jnl{PASP}}               
\def\pasj{\ref@jnl{PASJ}}               
\def\qjras{\ref@jnl{QJRAS}}             
\def\skytel{\ref@jnl{S\&T}}             
\def\solphys{\ref@jnl{Sol.~Phys.}}      
\def\sovast{\ref@jnl{Soviet~Ast.}}      
\def\ssr{\ref@jnl{Space~Sci.~Rev.}}     
\def\zap{\ref@jnl{ZAp}}                 
\def\nat{\ref@jnl{Nature}}              
\def\iaucirc{\ref@jnl{IAU~Circ.}}       
\def\aplett{\ref@jnl{Astrophys.~Lett.}} 
\def\apspr{\ref@jnl{Astrophys.~Space~Phys.~Res.}}
\def\bain{\ref@jnl{Bull.~Astron.~Inst.~Netherlands}}
\def\fcp{\ref@jnl{Fund.~Cosmic~Phys.}}  
\def\gca{\ref@jnl{Geochim.~Cosmochim.~Acta}}   
\def\grl{\ref@jnl{Geophys.~Res.~Lett.}} 
\def\jcp{\ref@jnl{J.~Chem.~Phys.}}      
\def\jgr{\ref@jnl{J.~Geophys.~Res.}}    
\def\jqsrt{\ref@jnl{J.~Quant.~Spec.~Radiat.~Transf.}}
\def\memsai{\ref@jnl{Mem.~Soc.~Astron.~Italiana}}
\def\nphysa{\ref@jnl{Nucl.~Phys.~A}}   
\def\physrep{\ref@jnl{Phys.~Rep.}}   
\def\physscr{\ref@jnl{Phys.~Scr}}   
\def\planss{\ref@jnl{Planet.~Space~Sci.}}   
\def\procspie{\ref@jnl{Proc.~SPIE}}   

\makeatother

\title[High resolution X-ray spectroscopy and imaging of Mrk~573]{High resolution X-ray spectroscopy and imaging of Mrk~573}

\author[Stefano Bianchi, et al.]{Stefano Bianchi$^1$\thanks{E-mail: bianchi@fis.uniroma3.it (SB)},  Marco Chiaberge$^{2,3}$, Daniel A. Evans$^4$,  Matteo Guainazzi$^5$ \newauthor Ranieri D. Baldi$^6$, Giorgio Matt$^1$, Enrico Piconcelli$^7$\\
$^1$Dipartimento di Fisica, Universit\`a degli Studi Roma Tre, via della Vasca Navale 84, 00146 Roma, Italy\\
$^2$Space Telescope Science Institute, 3700 San Martin Drive, Baltimore, MD 21218\\
$^3$INAF-IRA, Via P. Gobetti 101, I-40129 Bologna, Italy\\
$^4$MIT Kavli Institute for Astrophysics and Space Research / Harvard University\\
$^5$XMM-Newton Science Operations Center, European Space Astronomy Center, ESA, Apartado 50727, E-28080 Madrid, Spain\\
$^6$Universit\`a di Torino, via P. Giuria 1, 10125 Torino, Italy\\
$^7$Osservatorio Astronomico di Roma (INAF), Via Frascati 33, I-00040 Monte Porzio Catone, Italy\\
}

\begin{document}


\maketitle

\label{firstpage}

\begin{abstract}
We present a detailed analysis of the XMM-\textit{Newton} RGS high resolution X-ray spectra of the Seyfert 2 galaxy, Mrk~573. This analysis is complemented by the study of the \textit{Chandra} image, and its comparison to optical (\textit{HST}) and radio (\textit{VLA}) data. The soft X-ray emission is mainly due to gas photoionised by the central AGN, as indicated by the detection of radiative recombination continua from {O\,\textsc{vii}} and {O\,\textsc{viii}}, as well as by the prominence of the {O\,\textsc{vii}} forbidden line. This result is confirmed by the best fit obtained with a self-consistent \textsc{cloudy} photoionisation model. However, a collisionally excited component  is also required, in order to reproduce the {Fe\,\textsc{xvii}} lines, accounting for about 1/3 of the total luminosity in the 15-26 \AA\ band. Once adopted the same model in the \textit{Chandra} ACIS data, another photoionised component, with higher ionisation parameter, is needed to take into account emission from higher Z metals. The broadband ACIS spectrum also confirms the Compton-thick nature of the source. The imaging analysis shows the close morphological correspondence between the soft X-ray and the [{O\,\textsc{iii}}] emission. The radio emission appears much more compact, although clearly aligned with the narrow line region. The collisional phase of the soft X-ray emission may be due to starburst, requiring a star formation rate of $\simeq5-9$ M$_\odot$ yr$^{-1}$, but there is no clear evidence of this kind of activity from other wavelengths. On the other hand, it may be related to the radio ejecta, responsible for the heating of the plasma interacting with the outflow, but the estimated pressure of the hot gas is much larger than the pressure of the radio jets, assuming equipartition and under reasonable physical parameters.
\end{abstract}


\section{Introduction}

Important progress has been made in the last few years to unveil the origin of the soft X-ray emission in obscured Active Galactic Nuclei (AGN). The first breakthrough was represented by high resolution spectra made available thanks to the gratings aboard \textit{Chandra} and XMM-\textit{Newton}. The `soft excess' observed in CCD spectra was found to be due to the blending of bright emission lines, mainly from He- and H-like transitions of light metals and L transitions of Fe, with low or no continuum, in most Seyfert 2 galaxies \citep[see e.g.][]{sako00b,Sambruna01b,kin02,gb07}. Spectral diagnostic tools agree that the observed lines should be produced in a gas photoionised by the AGN, with little contribution from any collisionally ionised plasma. A second breakthrough was made possible thanks to the unrivaled spatial resolution of \textit{Chandra}. The soft X-ray emission of Seyfert 2 galaxies appears to be morphologically correlated with that of the Narrow Line Region (NLR), as mapped by the [{O\,\textsc{iii}}] $\lambda 5007$ \textit{HST} images \citep[e.g.][]{yws01,iwa03,bianchi06}. Since the NLR is also believed to be a gas photoionised by the AGN, it was shown that a very simple model where the soft X-ray emission and the NLR emission are produced in the same material is possible \citep[e.g.][]{bianchi06}.

However, this scenario is clearly oversimplified. In particular, it is not clear whether different components of the same medium are spatially separated, possibly radially distributed, or co-exists at each radius, as a result of stratification. Moreover, the role of radio ejecta, which are often found to be strongly correlated with the morphology of the NLR, is still unclear. An exciting possibility to shed some more light on this issue is provided by spatially resolved X-ray spectroscopy, which is, unfortunately, limited by the compact structures of this class of objects (at most some arcsec). Indeed, such a study was performed with high resolution spectroscopy on only one source, NGC~1068, being very bright and extended \citep{brink02}. The results are in agreement with the expectations from a cone of plasma, irradiated by the central AGN. On the other hand, a similar analysis, but with CCD resolution, was performed on NGC~7582, showing that there are regions with a further source of ionisation, or lower density \citep{bianchi07b}.

Mrk~573 (a.k.a. UGC~1214, z=0.0172) is the third [{O\,\textsc{iii}}] brightest source in the \citet{schm03} sample of nearby Seyfert galaxies observed by \textit{HST}, only fainter than NGC~1068 and Mrk~3. Moreover, the NLR extension of Mrk~573 is by far the largest of the sample, reaching a total extent of 3 kpc, with a projected extension almost 9 arcsec wide. A triple radio source is associated to the galaxy, composed by a central core and two spots \citep{uw84}. A detailed analysis of the NLR of this source was performed, among others, by \citet{ferr99} and \citet{schl09}. They conclude that photoionisation by the central AGN is likely the dominant process, although the interaction with the radio jets must be taken into account in the overall scenario, introducing kinematic disturbances and shaping the NLR morphology.

Despite the brightness and the interesting features of Mrk~573, the source has never been studied in detail in X-rays. After the detection by \textit{Einstein}, Mrk~573 was observed by \textit{ROSAT}, which clearly detected an emission extended over about 10 arcsec, in agreement with the NLR dimensions. The source was then only observed by XMM-\textit{Newton} with a short exposure time of about 10 ks. It confirmed to be rather bright in the 0.5-2 keV band ($\simeq3\times10^{-13}$ cgs) and the detection of a very strong iron line, with an EW larger than 1 keV, revealed its nature as a Compton-thick source \citep{gua05b}. But the most important piece of information comes from the RGS high resolution soft X-ray spectrum, which clearly appears dominated by strong emission lines. The predominance of K lines, the ratio of the components of the {O\,\textsc{vii}} triplet and the detection of strong, narrow, radiative recombination continua (RRC) features all contribute to an interpretation in terms of photoionised gas \citep{gb07}.

In this paper, we re-analysed in detail the XMM-\textit{Newton} RGS spectrum of Mrk~573, adopting a self-consistent photoionisation model. Moreover, we present for the first time the imaging and spectral analysis of a \textit{Chandra} observation.

\section{Observations and data reduction}

In the following, errors correspond to the 90 per cent confidence level for one interesting parameter ($\Delta \chi^2 =2.71$), where not otherwise stated. The adopted cosmological parameters are $H_0=70$ km s$^{-1}$ Mpc$^{-1}$, $\Omega_\Lambda=0.73$ and $\Omega_m=0.27$ \citep[i.e. the default ones in \textsc{xspec 12.5.1}:][]{xspec}. At the distance of Mrk~573, 1 arcsec corresponds to 360 pc. In all the fits, the Galactic column density along the line of sight to Mrk~573 is included \citep[$\mathrm{N_H}=2.96\times10^{20}$ cm$^{-2}$:][]{dl90}.

\subsection{X-rays: \textit{Chandra} and \textit{XMM-Newton}}

Mrk~573 was observed by \textit{Chandra} on 2006-11-18 for a total exposure time of 40 ks (obsid 7745), with the Advanced CCD Imaging Spectrometer \citep[ACIS:][]{acis}. Data were reduced with the Chandra Interactive Analysis of Observations \citep[CIAO:][]{ciao} 4.1 and the Chandra Calibration Data Base (CALDB) 4.1.2 software, adopting standard procedures. Images were corrected for known aspect offsets, reaching a nominal astrometric accuracy of 0.6 arcsec (at the 90 per cent confidence level). Spectra were extracted from three different regions: a circular region of 16 arcsec of radius (the default one analysed in Sect.~\ref{chandrafit}); the same region, but excluding the inner 1 arcsec; the nuclear region, i.e. the inner 1 arcsec. The imaging analysis was performed on event files without the pixel randomization and treated with the SER procedure \citep{li03}, in order to improve the positional accuracy. This allowed us to use a pixel size of 0.246 arcsec.

Mrk~573 was also observed by XMM-\textit{Newton} on 2004-01-15 for a total exposure time of $\mathbf{\simeq12}$ ks (obsid 0200430701), with the EPIC CCD cameras, the pn \citep{struder01} and two MOS \citep{turner01}, operated in Prime Full Window and Medium Filter. These data were already presented by \citet{gua05b} and \citet{gb07}. In this paper, we present a new and more detailed analysis of the RGS spectra, which were extracted with standard procedures, using SAS 8.0.1 \citep[last described in][]{sas610} and the most updated calibration files available at the time the data reduction was performed (October 2009). Background spectra were generated using blank field event lists, accumulated from different positions on the sky vault along the mission.

\begin{figure}
\begin{center}
\epsfig{file=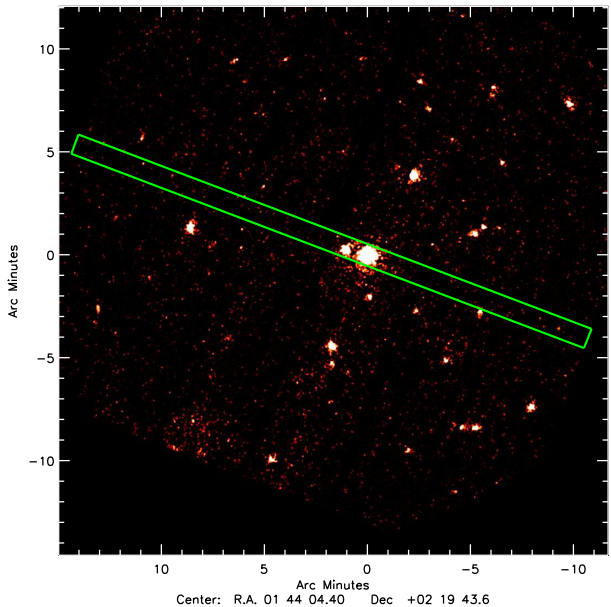, width=0.95\columnwidth}
\end{center}
\caption{\label{mrk573field}XMM-\textit{Newton} EPIC pn image (0.3-2 keV). The RGS extraction region is shown in green.}
\end{figure}

As shown in Fig.~\ref{mrk573field}, in the field there are several bright sources in the soft X-ray band. However, the extraction region (corresponding to the default 90 per cent of the point spread function in the cross-dispersion direction) includes only one contaminating source, named [TUM93]~J014401.5+022106 \citep{turner93}. We extracted its pn spectrum, and calculated a 0.5-0.8 keV flux of $1.6\times10^{-14}$ erg cm$^{-2}$ s$^{-1}$, i.e. a factor of 8 less than Mrk~573. It is unlikely to contaminate the line emission spectrum of Mrk~573, since any line, if present, would be displaced by the order of 1 \AA\ (without considering the effect of its unknown redshift), given its distance of around 1 arcmin from the nucleus. Such lines would therefore be at unidentified wavelengths, but none of them are detected (see Sect.~\ref{rgsanalysis}).

We also extracted the pn spectrum of Mrk~573, already analysed by \citet{gua05b}. However, since the quality of the data of this short observation is lower than that of the \textit{Chandra} ACIS spectrum, we will not discuss it further in this paper.

\subsection{Radio: VLA}

We downloaded from the \textit{VLA} archive the observation of Mrk~573 performed on 1985-03-03 at 6 cm (4860 MHz) in A array configuration. Data reduction was performed with the NRAO software package \textsc{AIPS}, following standard procedures, and very little flagging of bad data points. The phases of the calibrator 0146+056 were interpolated and applied to the object, and 3C~48 was used as the primary flux calibrator. The final map was produced using the \textsc{AIPS} task \textsc{IMAGR} with a beam size of 0$\farcs$5$\times$0$\farcs$4, cleaning depths of several thousand iterations and a CLEAN gain factor of 0.1.

\subsection{\label{hst}Optical and IR: HST}

We  retrieved the  \textit{HST}  observations we  use  in this  paper from  the Multimission  Archive  at STScI  (MAST).   The  images were  processed through  the  standard on-the-fly  reprocessing  system.  Mrk~573  was observed with \textit{HST} with WFPC2 and the FR533N ramp filter as part of the GO  program 6332 on  12/11/1995.  The  target was  located in  the WF2 camera to image the [{O\,\textsc{iii}}]$\lambda$5007 emission line.  The total exposure time
for the two CR-SPLITted exposures  is 600s.  We combine the two images using  the {\it  crrej} task  on {\it  IRAF} to  allow  for cosmic-ray rejection. The  image is clearly  dominated by the emission  line gas, therefore for  the purposes  of this work,  a careful  subtraction the continuum emission is not needed.  The object was also observed in the near-IR with  HST/NICMOS, as part  of GO7867.  Images were  taken with both  the F110W  and F160W  filters, which  are similar  to the  J and H-bands, respectively,  using the NIC1  camera.  The exposure  time is 1023s for both images. The data were originally presented by \citet{mp99}.  The  F110W filter  includes the  Pa$\beta$  emission line, while the F160W filter is relatively free of strong emission lines, at the redshift of Mrk~573. The image at longer wavelengths is clearly
dominated  by the  unresolved AGN  emission.  In  order to  assess the possible presence of structures in the central regions  of the galaxy (e.g. starburst regions, dust), we subtract the stellar emission of the host galaxy using  a model obtained by fitting  ellipses to the galaxy isophotes. The result  shows  that no  extra  nuclear structures  are
present. This image was therefore used as a reference for all the other images, by aligning the IR nucleus to the soft X-ray and [{O\,\textsc{iii}}] brightest pixels.

\section{Spectral analysis}

\subsection{\label{rgsanalysis}The soft X-ray RGS spectrum}

Following the procedure described in \citet{gb07}, the simultaneous fits on the spectra of the two RGS cameras were performed on $\simeq100$-channel-wide segments, adopting the Cash statistics \citep{cash76}. All the fits include a power law component ($\Gamma$ fixed to 1\footnote{Given the very limited band of these fits, the modellisation of the continuum is insensitive to the power law photon index, which can be frozen at any value.}) and as many emission lines as required (at the 90 per cent confidence level). A total of ten emission lines and two RRC were detected and identified securely (see Table \ref{rgslines}). The widths of all the features were consistent with being unresolved. The RRC are then properly modelled with the \textsc{redge} model, which fully takes into account the recombination edge profile, giving a direct estimate of the electron temperature of the gas.

\begin{table}
\caption{\label{rgslines}Mrk~573: detected emission lines and RRC in the XMM-\textit{Newton} RGS spectrum.}
\begin{center}
\begin{tabular}{lllll}
$\mathbf{E_{o}}$ & \textbf{Line id.} & $\mathbf{E_{th}}$ & kT & \textbf{Flux}\\
&&&\\
$0.5007\pm0.0011$ & {N\,\textsc{vii}} K$\alpha$ & 0.5003 &  -- & $1.3^{+1.2}_{-0.9}$\\
$0.5614\pm0.0004$ &{O\,\textsc{vii}} K$\alpha$ (f) & 0.5610 &  -- & $5.4^{+2.9}_{-2.1}$\\
$0.5739^{+0.0003}_{-0.0007}$ & {O\,\textsc{vii}} K$\alpha$ (r) & 0.5739 &  -- & $3.1^{+2.2}_{-1.5}$\\
$0.6541^{+0.0007}_{-0.0006}$ & {O\,\textsc{viii}} K$\alpha$ & 0.6536 &  -- & $2.2^{+1.2}_{-1.0}$\\
$0.6964^{+0.0015}_{-0.0010}$ & {O\,\textsc{vii}} K$\gamma$ & 0.6978 &  -- & $0.9^{+0.7}_{-0.5}$\\
\multirow{2}{*}{$0.7266^{+0.0013}_{-0.0015}$} & {Fe\,\textsc{xvii}} M2 & 0.7252 &  \multirow{2}{*}{--} & \multirow{2}{*}{$1.2^{+0.8}_{-0.6}$}\\
& {Fe\,\textsc{xvii}} 3G & 0.7272 & &\\
\multirow{2}{*}{$0.7389\pm0.0015$} & {O\,\textsc{vii}} RRC & 0.7393 &  $<14$ & \multirow{2}{*}{$1.4^{+1.0}_{-0.7}$}\\
& {Fe\,\textsc{xvii}} 3F & 0.7390 & -- &\\
\multirow{2}{*}{$0.776^{+0.002}_{-0.003}$} & {O\,\textsc{viii}} K$\beta$ & 0.7746 &  \multirow{2}{*}{--} & \multirow{2}{*}{$0.6^{+0.6}_{-0.4}$}\\
& {Fe\,\textsc{xviii}} L & 0.7747 & &\\
$0.826\pm0.002$ & {Fe\,\textsc{xvii}} 3C & 0.8257 &  -- & $0.4^{+0.5}_{-0.3}$\\
\multirow{3}{*}{$0.8705^{+0.0017}_{-0.002}$} & {Fe\,\textsc{xviii}} L & 0.8626 & --  & \multirow{3}{*}{$1.0^{+0.7}_{-0.5}$}\\
& {O\,\textsc{viii}} RRC & 0.8714 & $<15$ &\\
& {Fe\,\textsc{xviii}} L & 0.8728 & -- &\\
\end{tabular}
\end{center}

Energies are in units of keV, fluxes of $10^{-5}$ ph cm$^{-2}$ s$^{-1}$, kT in eV. Theoretical energies are from CHIANTI \citep{dere97,dere09}. The labelling for {Fe\,\textsc{xvii}} lines follows that of \citet{brown98}.
\end{table}

\begin{figure*}
\begin{center}
\epsfig{file=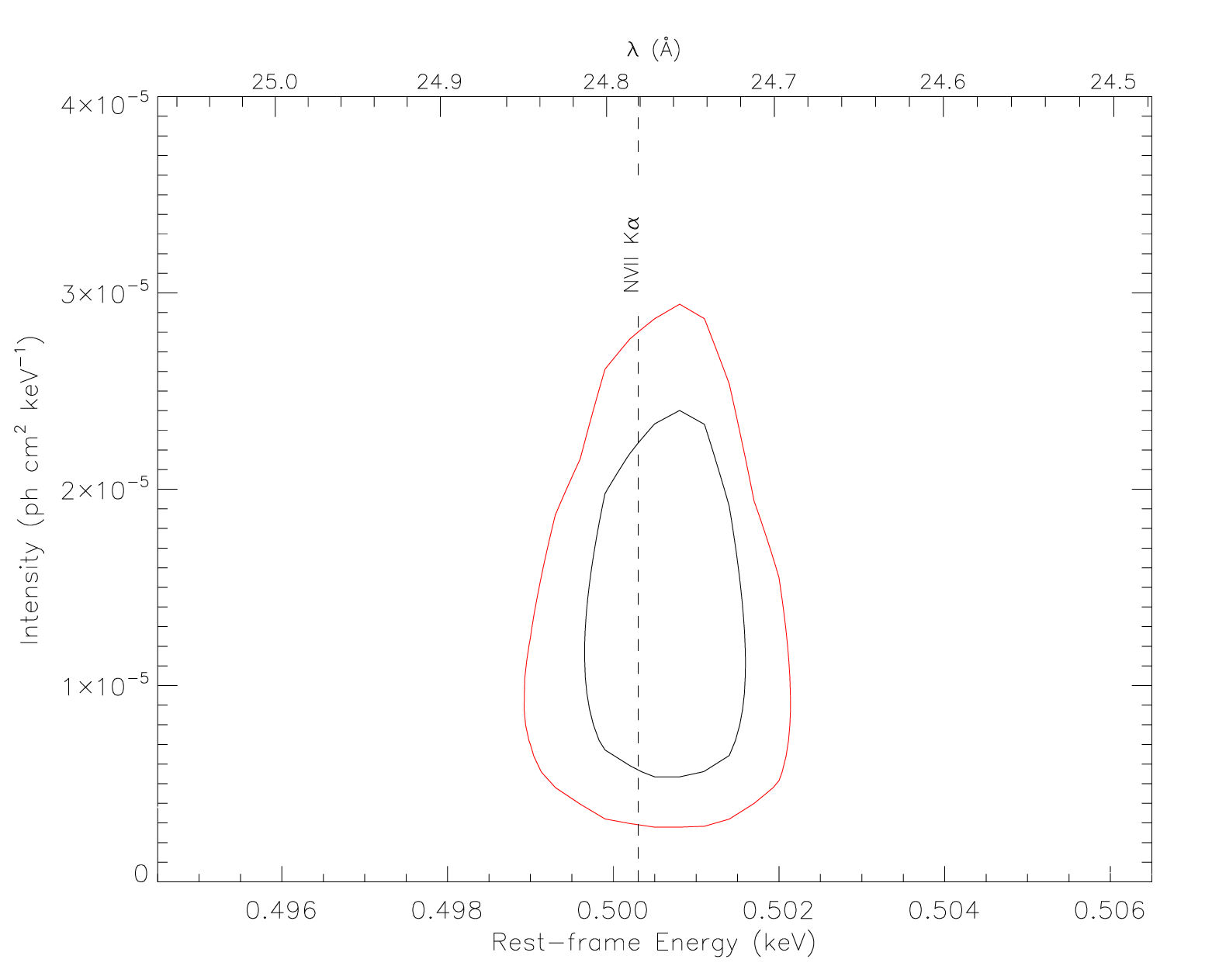, width=0.8\columnwidth}
\epsfig{file=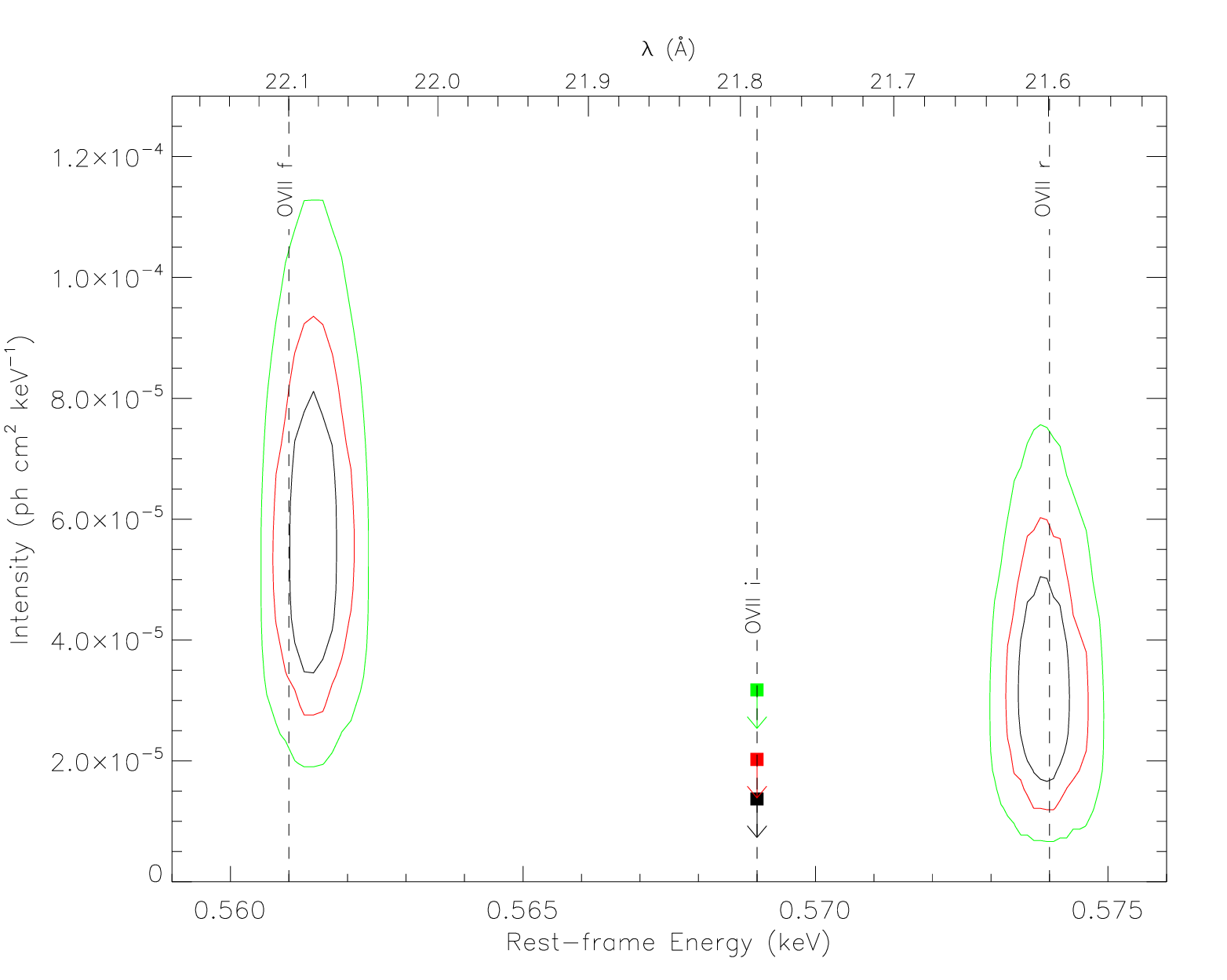, width=0.8\columnwidth}
\epsfig{file=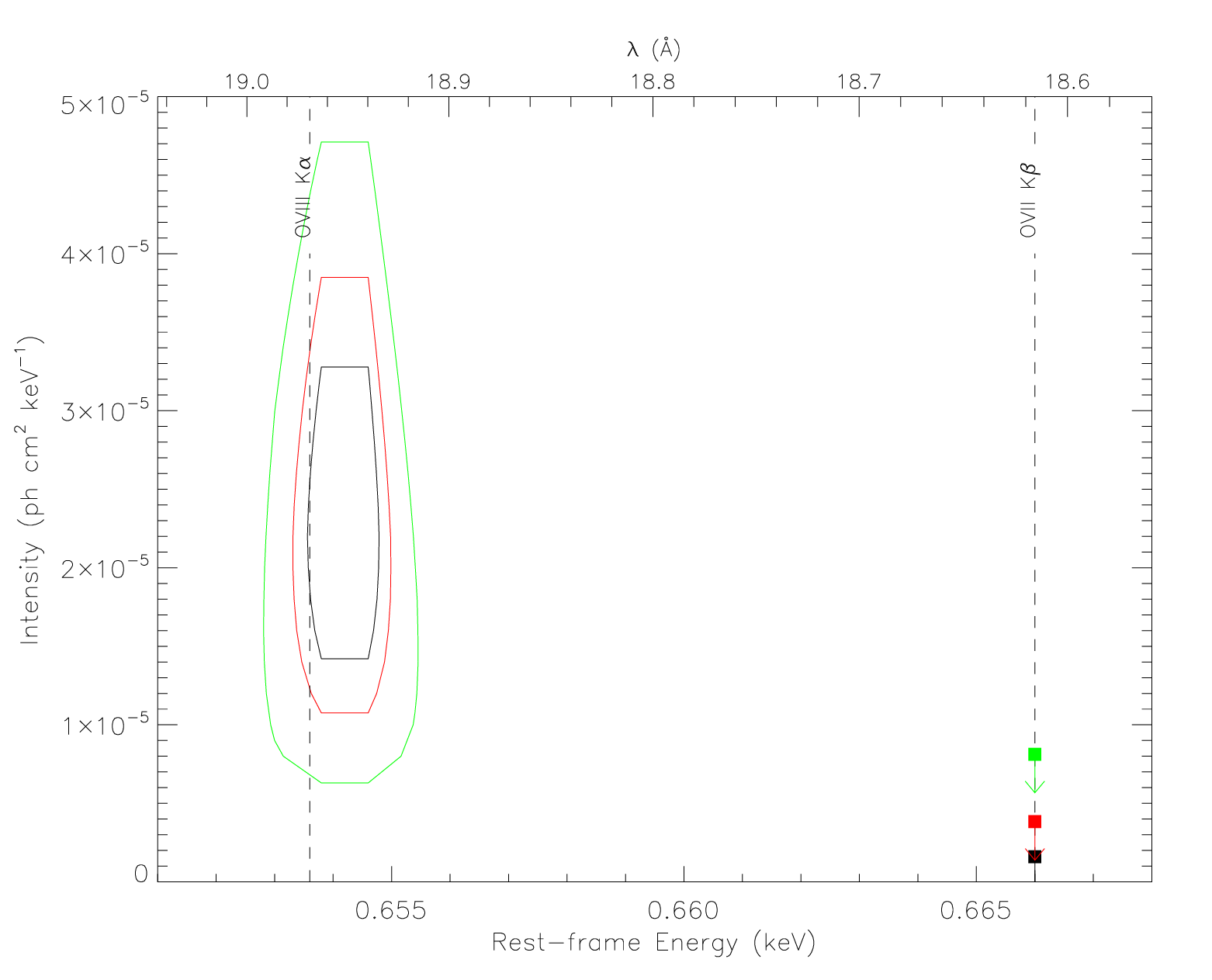, width=0.8\columnwidth}
\epsfig{file=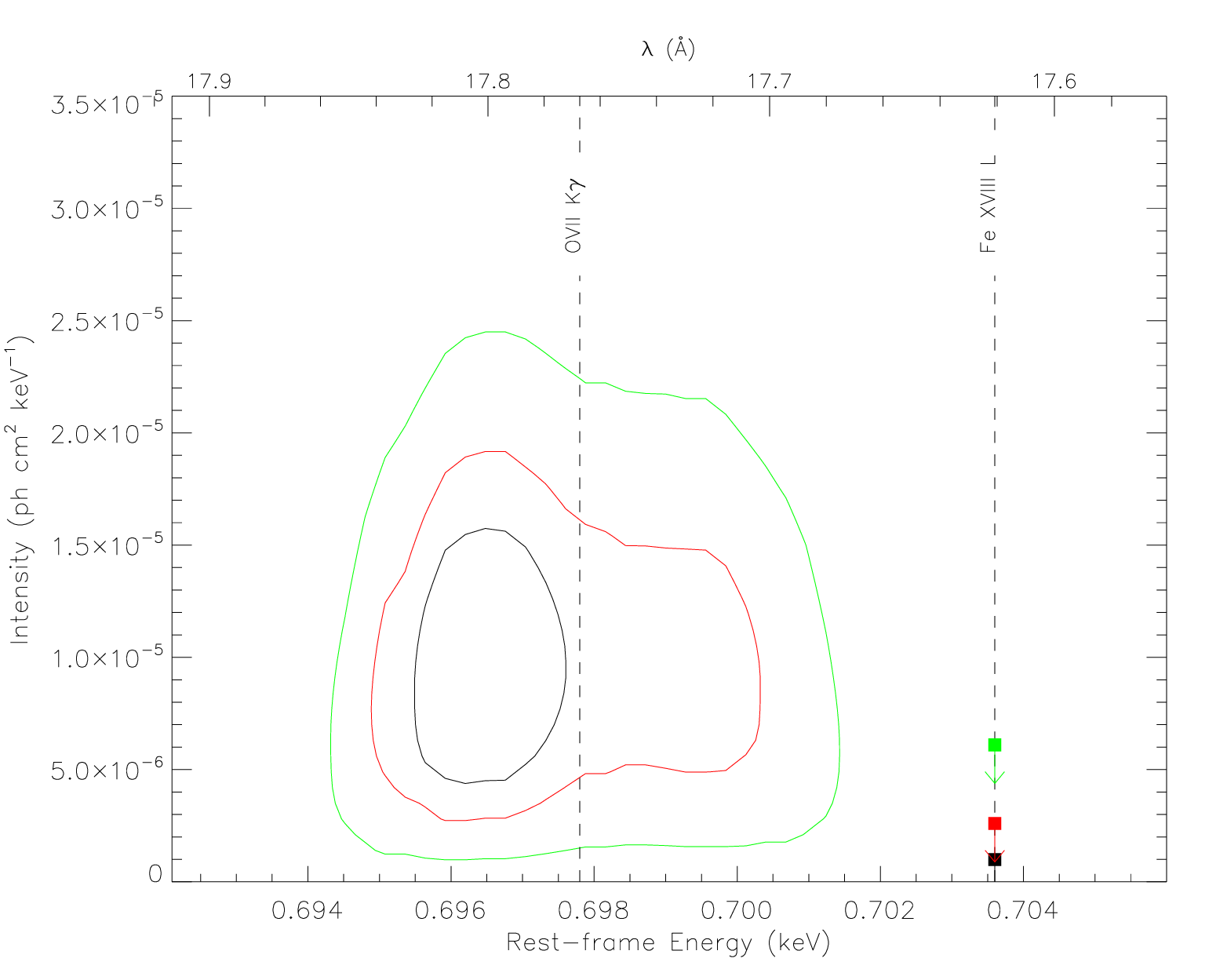, width=0.8\columnwidth}
\epsfig{file=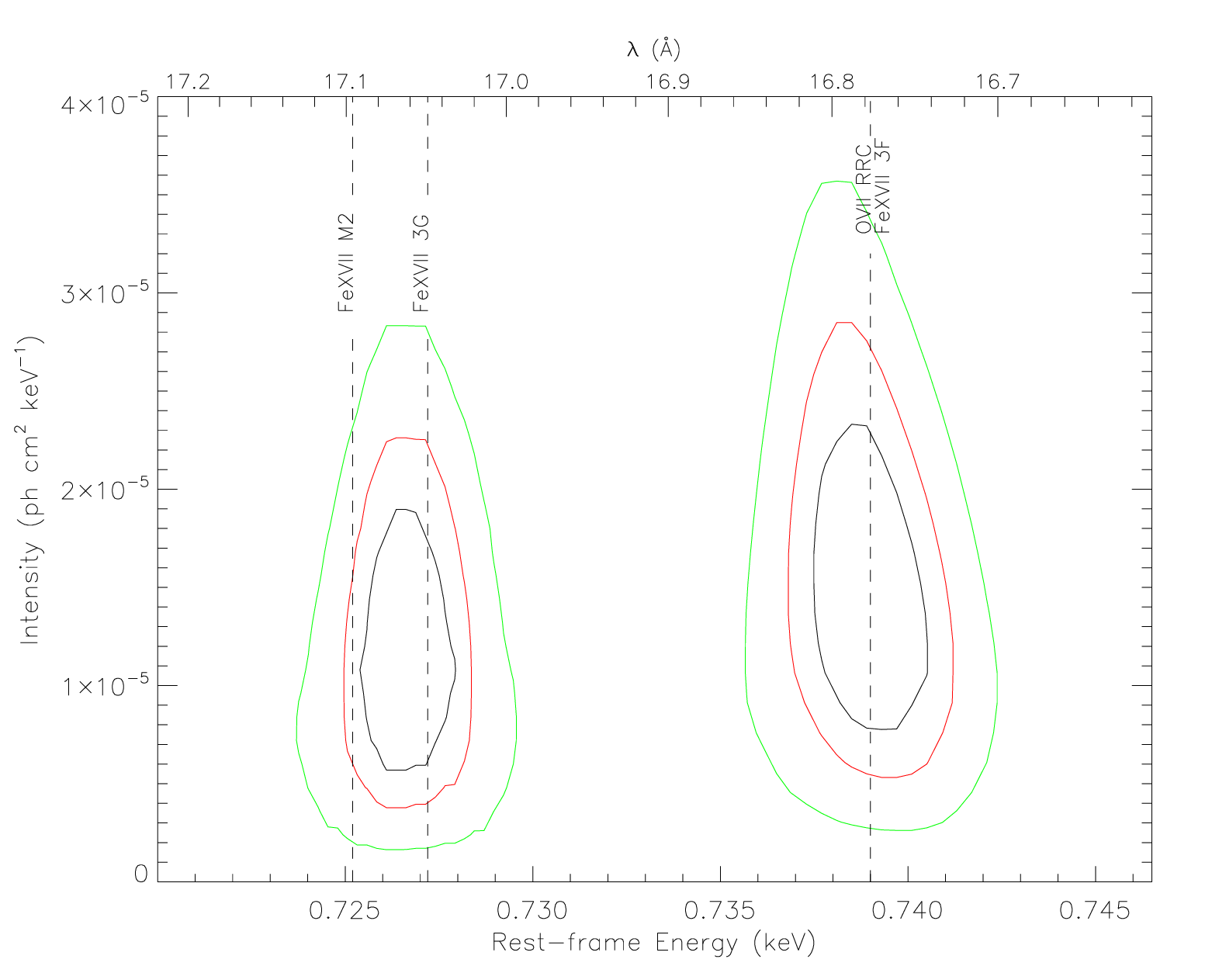, width=0.8\columnwidth}
\epsfig{file=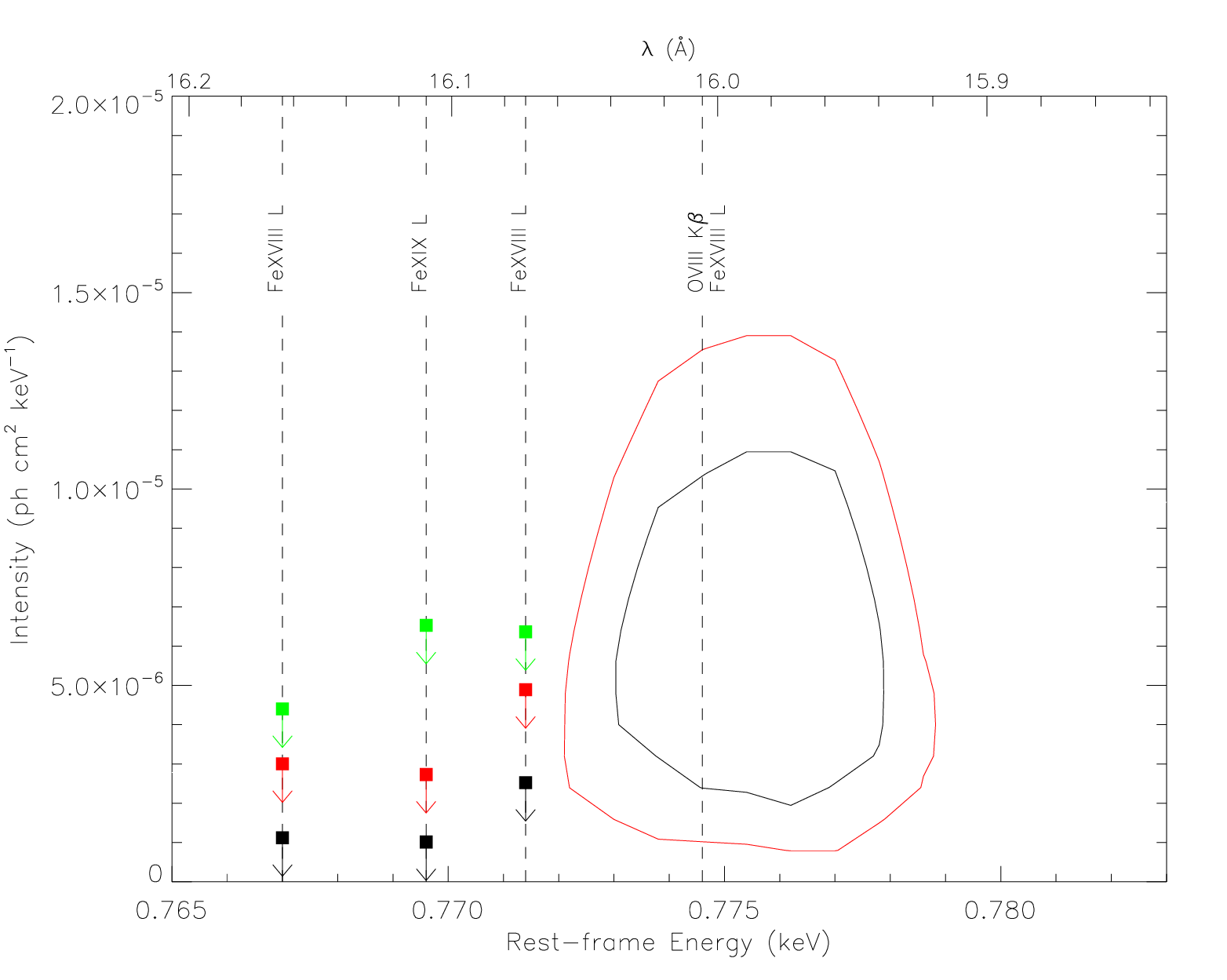, width=0.8\columnwidth}
\epsfig{file=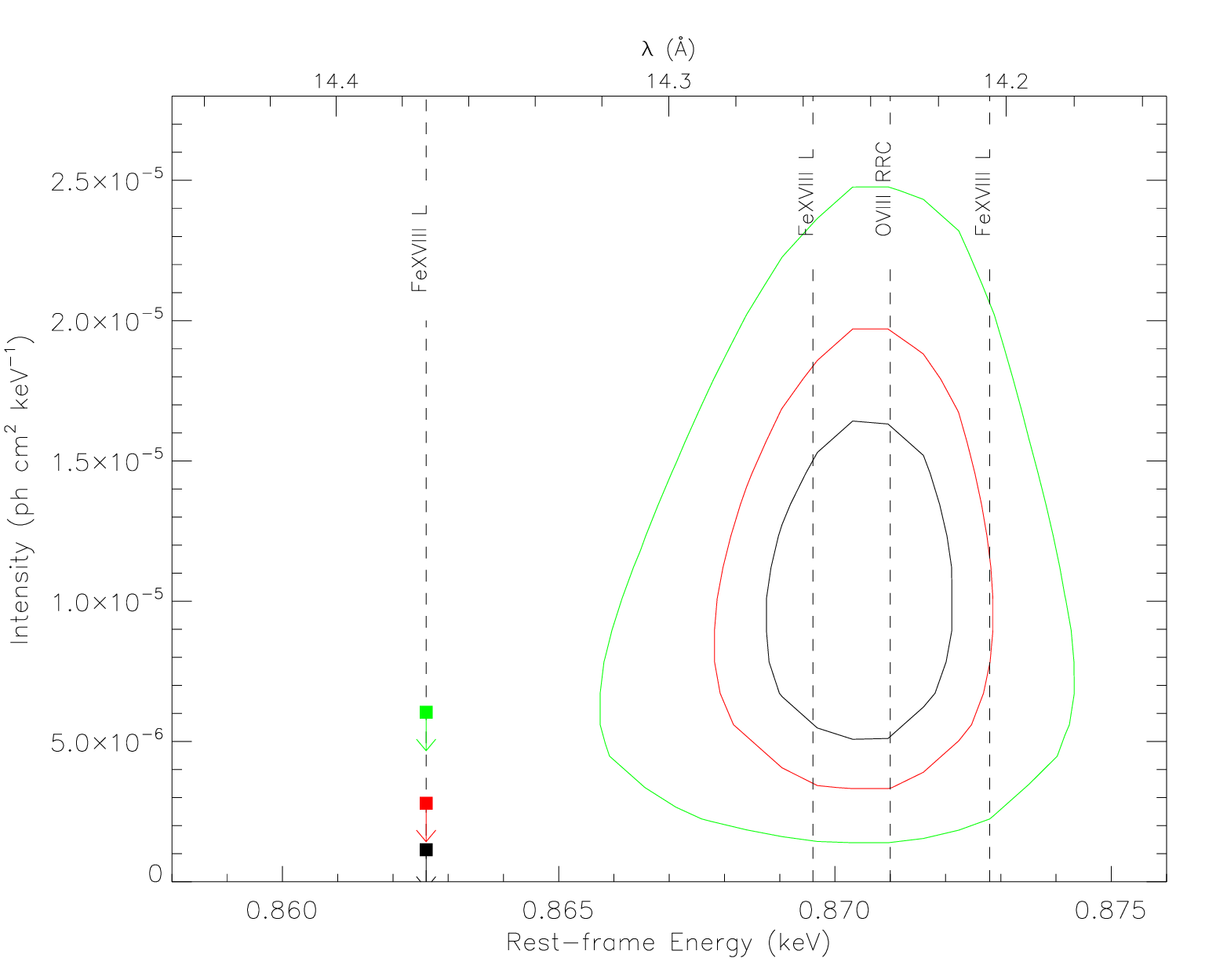, width=0.8\columnwidth}
\end{center}
\caption{\label{rgs}Mrk~573: centroid energy vs. flux contour plots for the detected emission lines in the XMM-\textit{Newton} RGS spectrum (see Table \ref{rgslines}). The contours refers to $\Delta$C=2.30, 4.61 and 9.21, i.e. confidence levels of 68, 90 and 99 per cent for two interesting parameters. Upper limits for the closest lines are also plotted ($\Delta$C=1.00, 2.71 and 4.00).}
\end{figure*}

The resulting temperature ($<15$ eV) is much less than would be expected from the same species, if the gas were in collisional equilibrium. This is a strong piece of evidence in favour of a dominant photoionised phase in the emitting region. However, the {O\,\textsc{vii}} RRC (0.7393 keV) may be significantly contaminated by the 3F component of {Fe\,\textsc{xvii}} L emission (0.7390 keV). This is suggested by the detection of the blend of the 3G and M2 components of the same species, at $\simeq0.7266$ keV. Our simulations with the \textsc{apec} model in \textsc{xspec} showed that the 3F/(3G+M2) ratio never exceeds $\simeq0.44$, for a wide range of temperatures. On the other hand, the observed ratio between the line detected at 0.7389 keV and that at 0.7266 keV is larger, being $1.2^{+0.7}_{-0.5}$. This means that a large part of the observed flux must be due to {O\,\textsc{vii}} RRC. The {O\,\textsc{viii}} RRC may also be contaminated by {Fe\,\textsc{xviii}} L lines (see Table~\ref{rgs}), but no other emission lines from that ion are detected in the spectrum.

In principle, diagnostics on the {O\,\textsc{vii}} triplet may also be a good indicator of the ionisation mechanism of the gas. Indeed, the predominance of the forbidden component is a sign of photoionisation, and the significant detection of the resonant transition a hint that pure recombination is not the only mechanism to produce the emission lines, but photoexcitation has also an important role. This, in turns, suggests that the gas column density should be not too large, in order not to suppress this process. However, given the large uncertainties on the line fluxes, it is impossible to exclude other solutions only on these grounds, like a contribution to the resonant line from a collisionally excited plasma.

Another hint for photoionisation comes from the {O\,\textsc{vii}} K$\alpha$ forbidden line to {O\,\textsc{viii}} K$\alpha$  ratio, which is larger than 1 ($2.5^{+1.9}_{-1.5}$). Together with a large total luminosity of the K$\alpha$ oxygen lines of $6\times10^{40}$ erg s$^{-1}$, this puts Mrk~573 in the photoionisation-dominated locus of the empirical diagnostic plot presented in \citet{gua09}.

Since the phenomenological analysis of the RGS spectrum favours an origin in a photoionised gas, we tried a more physical approach, performing a self-consistent fit on the whole spectrum (limited by the S/N to the $15-26$ \AA\ band).
We produced a grid model for \textsc{xspec} using \textsc{cloudy} 08.00 \citep[last described by][]{cloudy}. The main ingredients are: plane parallel geometry, with the flux of photons striking the illuminated face of the cloud given in terms of ionisation parameter $U$ \citep{of06}; incident continuum modelled as in \citet{korista97}; constant electron density $\mathrm{n_e}=10^5$ cm$^{-3}$; elemental abundances as in Table 9 of \textsc{cloudy} documentation\footnote{Hazy 1 version 08, p. 47: \url{http://viewvc.nublado.org/index.cgi/tags/release/c08.00/docs/hazy1_08.pdf?revision=2342&root=cloudy}}; grid parameters are $\log U=[-0.25:2.00]$, step 0.25, and $\log N_\mathrm{H}=[19.0:23.5]$, step 0.1. Only the reflected spectrum, arising from the illuminated face of the cloud, is taken into account in our model. We also produced tables with different densities ($\mathrm{n_e}=10^3-10^4$ cm$^{-3}$): all the fits presented in this paper resulted insensitive to this parameter, as expected since we are always treating density regimes where line ratios of He-like triplets are insensitive to density \citep{pd00}.

\begin{figure*}
\begin{center}
\epsfig{file=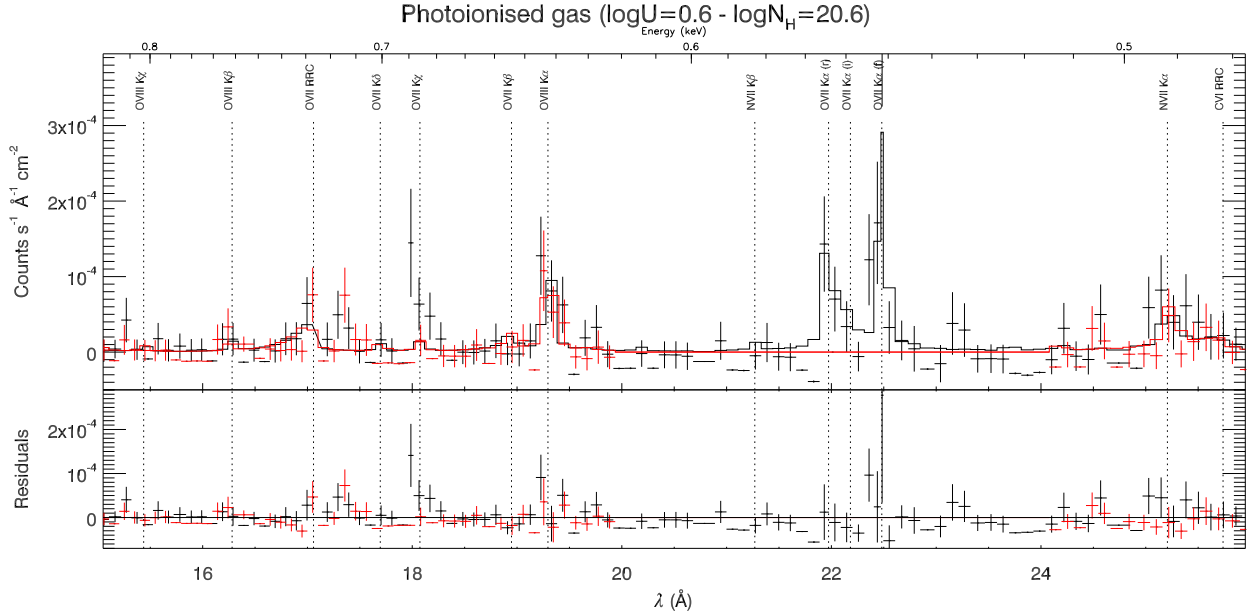,width=2\columnwidth}
\epsfig{file=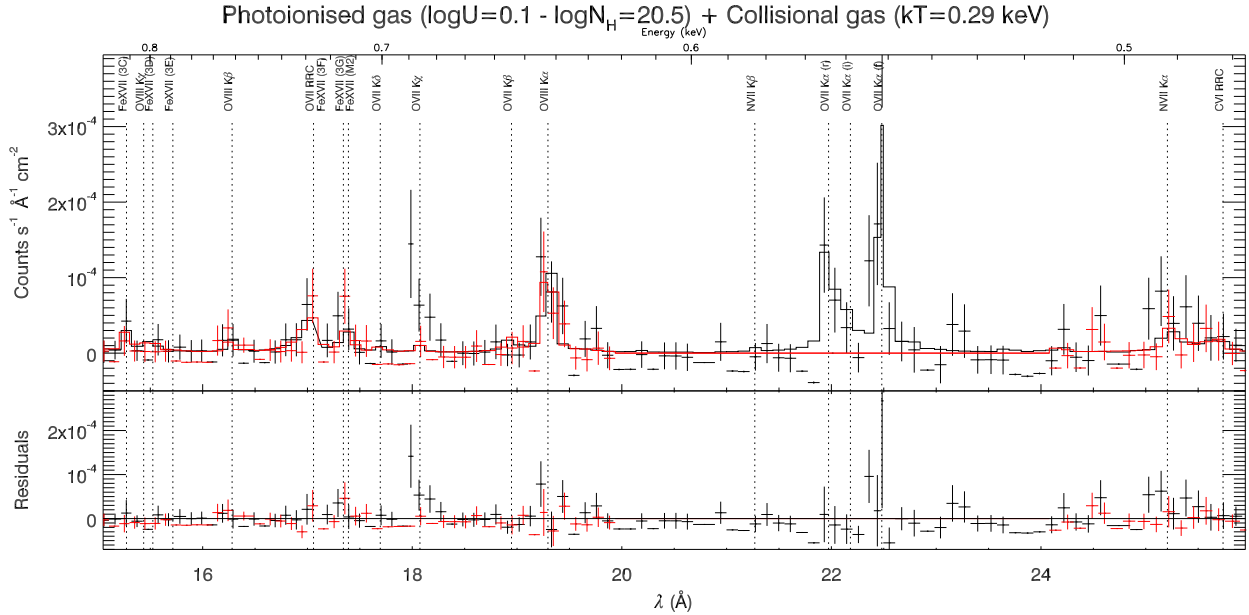,width=2\columnwidth}
\end{center}
\caption{\label{rgsfit}XMM-\textit{Newton} RGS spectra ($15-26$ \AA) of Mrk~573, rebinned for displaying purposes only. \textit{Top}: Best fit with a single photoionised phase, whose parameters are reported above. \textit{Bottom}: Best fit with a photoionised and a collisional phase, whose parameters are reported above. The brightest emission lines for the adopted models are labelled.}
\end{figure*}

The fit with a single photoionised phase is presented in the upper panel of Fig.~\ref{rgsfit}. With $\log\mathrm{U}=0.6\pm0.3$ and $\log\mathrm{N_H}=20.6\pm0.7$, a good fit is obtained, with most of the lines detected in our phenomenological analysis reasonably modelled, including, notably, the RRCs. However, there are clear residuals at the {Fe\,\textsc{xvii}} 3G+M2 wavelength: these lines are completely missing from the photoionised model\footnote{The \textsc{cloudy} line database was rather inaccurate for the {Fe\,\textsc{xvii}} transition wavelengths and relative atomic parameters. We therefore modified it according to the \textsc{chianti} database.}. This is not surprising, given the very low ionic fraction of {Fe\,\textsc{xvii}} ($\simeq4\times10^{-4}$) at this ionisation parameter.

We therefore tried to add another photoionised phase, with a larger U. However, any attempt to reproduce the observed {Fe\,\textsc{xvii}} lines largely overpredicts the {O\,\textsc{viii}} K$\alpha$ and K$\beta$ lines. This forced us to abandon this approach and to try an alternative scenario, where the iron L lines are mainly produced in a collisional gas. Indeed, the addiction of a collisional phase ($\mathrm{kT}=0.30^{+0.10}_{-0.06}$ keV) perfectly takes into account the {Fe\,\textsc{xvii}} 3G+M2 lines, together with the 3C component at 0.826 keV, without affecting too much the rest of the spectrum. The improvement of the Cash statistics for the addition of the collisional gas component is $\Delta C=18$, with two less degrees of freedom. No further improvement is achieved allowing the elemental abundances to vary. The parameters of the photoionised phase change to $\log\mathrm{U}=0.1^{+0.5}_{-0.7}$ and $\log\mathrm{N_H}=20.5^{+0.9}_{-1.0}$. The residuals of this hybrid (photoionised+collisional) gas are presented in the lower panel of Fig.~\ref{rgsfit} and are rather good. The only exception is represented by some positive residuals for the {O\,\textsc{vii}} K$\gamma$, but they are only significant in the RGS1 spectrum. Indeed, a separate analysis of the two RGS spectra leads to a line detection only in the RGS1 ($2.5\pm1.4\times10^{-5}$ ph cm$^{-2}$ s$^{-1}$ at $0.6995\pm0.0013$ keV, formally inconsistent with {O\,\textsc{vii}} K$\gamma$), while in the RGS2 only an upper limit is measured if the line energy is fixed at the theoretical one ($<0.8\times10^{-5}$ ph cm$^{-2}$ s$^{-1}$). We therefore conclude that the RGS1 detection is insecure, and therefore cannot draw any conclusions based on that only\footnote{We note here that the line wavelength does not correspond to any known defective pixel in neither of the RGS cameras (\url{http://xmm.esac.esa.int/external/xmm_user_support/documentation/uhb/node59.html#3177}).}.

The total flux in the 0.5-0.8 keV band is $1.3\times10^{-13}$ erg cm$^{-2}$ s$^{-1}$. The contribution of the collisional phase is $4.6\times10^{-14}$ erg cm$^{-2}$ s$^{-1}$, i.e. roughly 1/3 of the total flux in this band.

\subsection{\label{chandrafit}The \textit{Chandra} broadband spectrum}

The broadband (0.4-8 keV) \textit{Chandra} X-ray spectrum of Mrk~573 appears dominated by a strong emission in the soft band, plus strong neutral iron K$\alpha$ emission, typical signatures of an highly obscured AGN. The high energy part of the spectrum is well fitted by a pure neutral reflection component plus the iron line. The properties of the reflection component are unconstrained, due to the low statistics, so we fixed the photon index to 1.7 \citep[as in typical Seyfert galaxies, see e.g.][]{bianchi09} and the cosine of the inclination angle to 0.45. The largest normalization compatible with the data leads to an iron EW of $\simeq1.8$ keV, perfectly in agreement with the expectations for a Compton-thick AGN \citep{mbf96}. Since the normalization of the reflection component is basically unconstrained in the following fits, where the soft X-ray component may contribute also to the high energy part of the spectrum (see below), we decided to fix it. The 2-10 keV flux and the iron line flux are in perfect agreement with the values measured with XMM-\textit{Newton} \citep{gua05b}.

\begin{figure*}
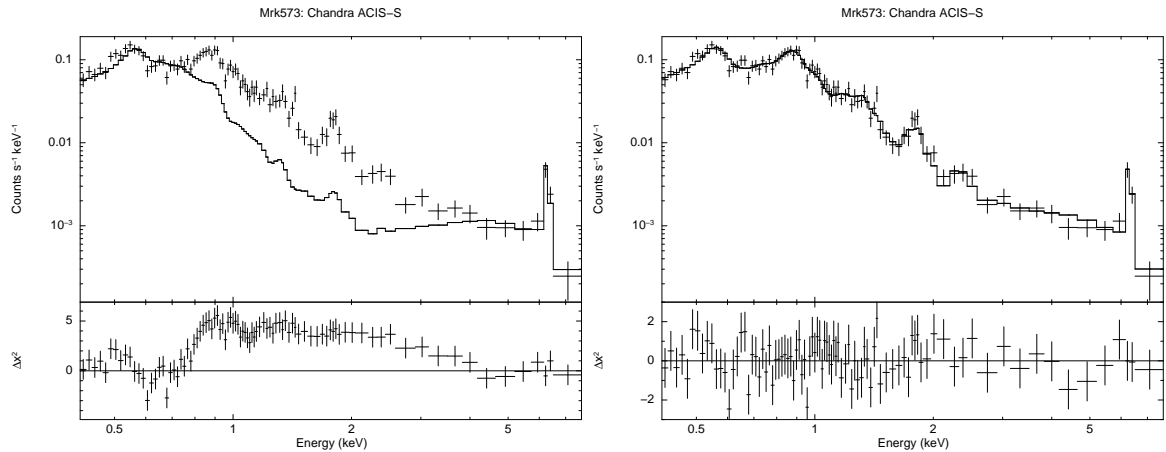

\begin{center}
\epsfig{file=mrk573_acis_rgsfit.ps, width=6cm, angle=-90}
\epsfig{file=mrk573_bestfit.ps, width=6cm, angle=-90}
\end{center}
\caption{\label{broadband}Left: Mrk~573 \textit{Chandra} ACIS-S broadband spectrum, with the best fit for the RGS data in the soft X-ray part, and $\Delta\chi^2$ residuals. Right: The same, but with the best fit.}
\end{figure*}

To fit the soft X-ray band, at first we adopted the same model which successfully reproduces the RGS spectrum. This model is a good representation of the ACIS spectrum, when limited to the same band, but clearly fails to fit the remaining part of the data, because it cannot reproduce the observed emission from higher Z metals (see left panel of Fig.~\ref{broadband}). We added another photoionised phase to the fit, with a larger ionisation parameter, and allowed also the parameters of the other phases to vary. The resulting fit is very good ($\chi^2=74/79$ d.o.f., see right panel Fig~\ref{broadband} and Table~\ref{broadbandfit}). The best fit model includes two further emission lines, which can be readily identified with neutral Si K$\alpha$ and S K$\alpha$ at 1.740 keV and 2.308 keV, respectively \citep{house69}, likely arising from the same Compton-thick material responsible for the production of the features which dominate the high energy part of the spectrum, that is the Compton reflection component and the neutral iron K$\alpha$ line.

The best fit parameters for the two photoionised phases are $\log\mathrm{U}=0.3^{+0.3}_{-0.6}$ and $\log\mathrm{N_H}=21.7^{+0.6}_{-0.8}$, and $\log\mathrm{U}=1.81^{+0.15}_{-0.12}$ and $\log\mathrm{N_H}=21.8^{+0.7}_{-0.8}$. The temperature of the collisional gas is $\mathrm{kT}=0.56\pm0.12$ keV.

If the three-phase model is re-applied to the RGS spectra, the fit is visually comparable to the one obtained above directly on these data, with only two phases. Therefore, although only 2 phases (one photoionised, the other collisional) is required by the RGS data, the \textit{Chandra} ACIS data reveals the presence of a further photoionised component, still allowed by the XMM-\textit{Newton} high-resolution spectrum.

The total 0.5-2 keV flux is $2.7\times10^{-13}$ erg cm$^{-2}$ s$^{-1}$, in agreement with the one measured with XMM-\textit{Newton} \citep{gua05b}. The contribution from the collisional phase is $\simeq20\%$, corresponding to an unabsorbed luminosity of $4.0\times10^{40}$ erg s$^{-1}$ in the same band. It is interesting to note that, of the remaining photoionised phases (which contribute roughly equally to the the soft X-ray flux), $\simeq15\%$ is constituted by a continuum component (which includes RRCs and the Thomson-scattered power law), while the remaining flux is in emission lines. 

\begin{table}
\caption{\label{broadbandfit}Best fit parameters for the \textit{Chandra} spectrum of Mrk~573. Fluxes are in units of $10^{-13}$ erg cm$^{-2}$ s$^{-1}$, line fluxes in units of $10^{-6}$ ph cm$^{-2}$ s$^{-1}$, energies and kT in keV. See text for details.}
\begin{center}
\begin{tabular}{ll}
 &  \\
$\log\mathrm{U_1}$&  $0.3^{+0.3}_{-0.6}$\\
$\log\mathrm{N_{H1}}$ & $21.7^{+0.6}_{-0.8}$ \\
$\log\mathrm{U_2}$ & $1.81^{+0.15}_{-0.12}$ \\
$\log\mathrm{N_{H2}}$ & $21.8^{+0.7}_{-0.8}$ \\
kT & $0.56\pm0.12$\\
E$_\mathrm{Si K\alpha}$ & 1.740$^*$ \\
F$_\mathrm{Si K\alpha}$ & $1.0\pm0.7$ \\
E$_\mathrm{S K\alpha}$ & 2.308$^*$ \\
F$_\mathrm{S K\alpha}$ & $1.2\pm0.8$ \\
E$_\mathrm{Fe K\alpha}$ & $6.35\pm0.03$ \\
F$_\mathrm{Fe K\alpha}$ & $5.5\pm1.7$ \\
$\chi^2$/dof & 74/79 \\
  &   \\
$F_{0.5-2}$ & $2.7\pm0.3$ \\
$F_{2-10}$ & $2.8\pm0.5$ \\
  &  \\
\end{tabular}
\end{center}
$^*$ fixed
\end{table}

\section{Imaging analysis}

\begin{figure*}
\begin{center}
\epsfig{file=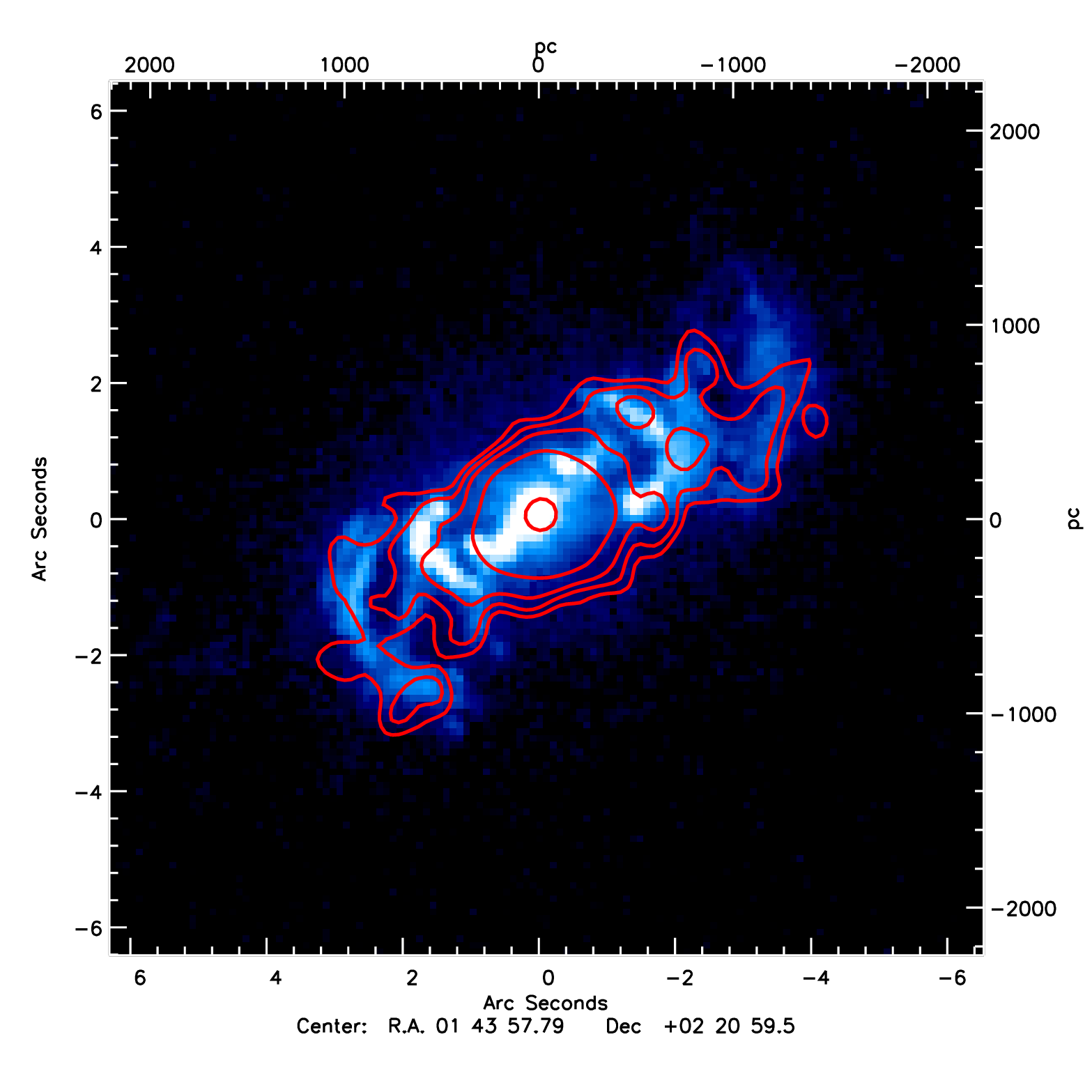, width=\columnwidth}
\epsfig{file=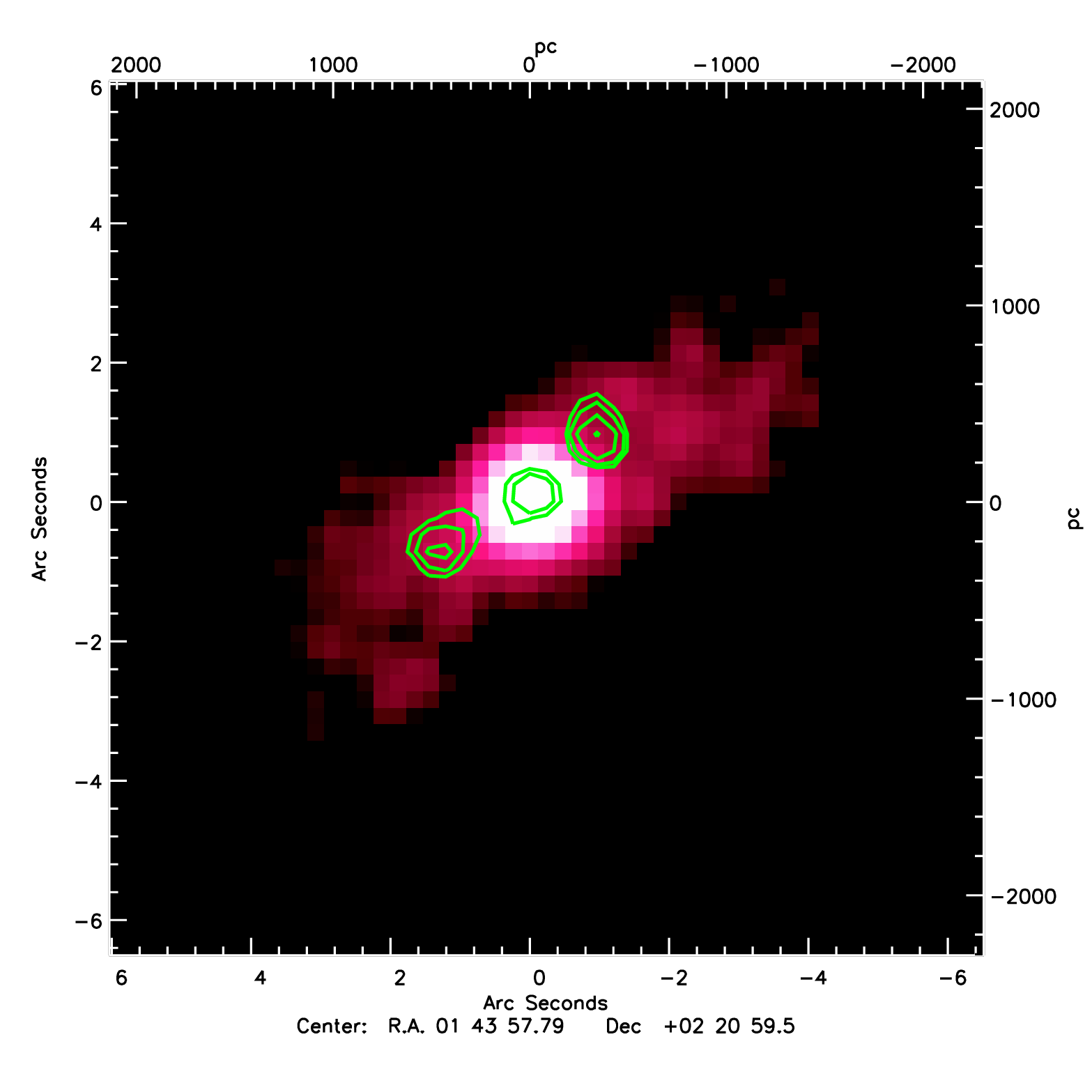, width=\columnwidth}
\end{center}
\caption{\label{o3_softx}Left: \textit{Chandra} soft X-ray (0.2-2 keV) contours superimposed on the \textit{HST} {O\,\textsc{iii}} image. The contours refers to 0.001, 0.1, 0.2, 0.4 and 0.95 levels with respect to the brightest pixel. North is up, east to the left. Right: VLA radio (6 cm) contours superimposed on the \textit{Chandra} soft X-ray (0.2-2 keV) image.}
\end{figure*}

The soft X-ray (0.2-2 keV) emission of Mrk~573 is clearly extended and closely resembles the morphology of the NLR, as mapped by the [{O\,\textsc{iii}}] \textit{HST} image (see left panel of Fig.~\ref{o3_softx}). This is often observed in Seyfert 2 galaxies, and clearly suggests a common physical origin for the two emissions \citep[e.g.][]{bianchi06}. Given the lower angular resolution of \textit{Chandra}, it is hard to say how good the correspondence between the X-rays and the NLR is at the smallest scale. However, there is no evidence that significant deviations are present in these data.

On the other hand, the right panel of Fig~\ref{o3_softx} shows the \textit{VLA} radio emission superimposed on the \textit{Chandra} data. As already reported by \citet{uw84} and \citet{fal98}, Mrk~573 has a unresolved radio core of a diameter size of $\sim$0.32 kpc (1 arcsec) and two-sided jets: the northwest and the southeast bubbles have a similar size of $\sim$0.48$\times$0.32 kpc$^{2}$ (1$\farcs$5$\times$1$\farcs$0). Table \ref{radiotable} reports flux and luminosities for these three radio components. It is clear that, although the inclination of the radio jets is aligned with that of the soft X-ray emission, the overall extension of the radio emission is far more compact.

\begin{table}
\caption{\label{radiotable}Flux (mJy) and luminosity (erg s$^{-1}$ Hz$^{-1}$) at 6 cm for the three radio components detected in Mrk~573.}
\begin{center}
\begin{tabular}{lll}
& &  \\
& Flux & $\mathrm{\log L}$ \\
core & $0.95\pm0.03$ &  27.72\\
NW & $2.22\pm0.03$ & 28.09 \\
SE & $4.54\pm0.05$ & 28.40 \\
& & \\
\end{tabular}
\end{center}
\end{table}

\begin{figure*}
\begin{center}
\epsfig{file=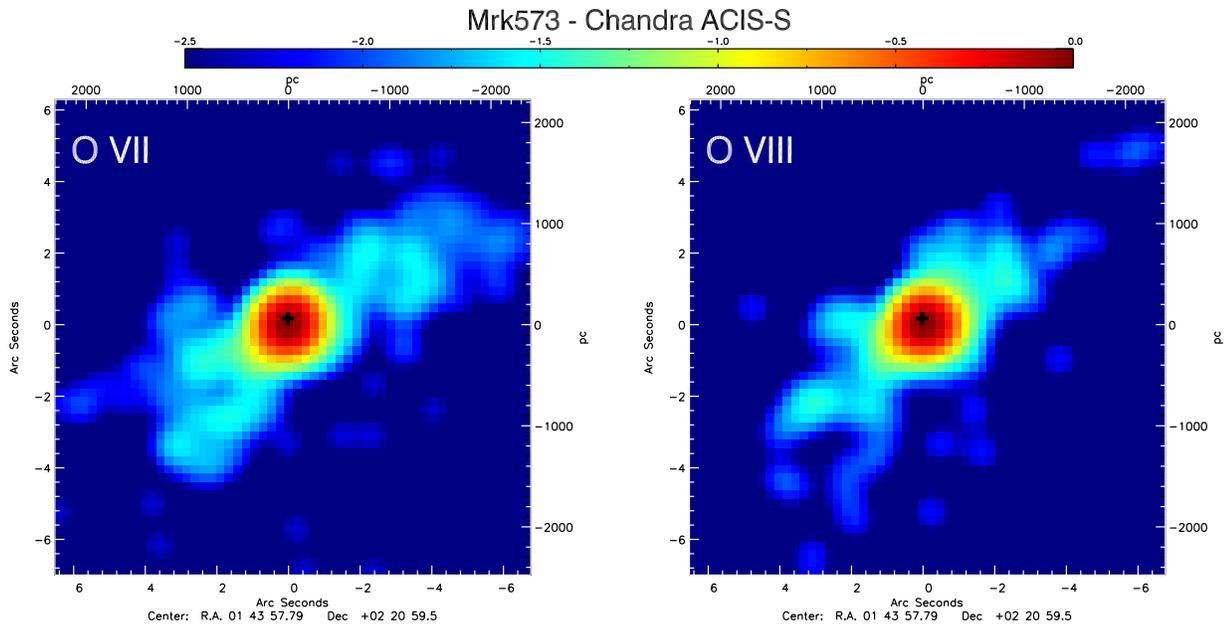, width=2\columnwidth}
\end{center}
\caption{\label{o7_o8}Mrk~573: \textit{Chandra} images in the 0.5-0.6 keV ({O\,\textsc{vii}}) and 0.6-0.7 keV ({O\,\textsc{viii}}) band. The images were smoothed with a 3x3 FWHM Gaussian filter and normalised to the brightest pixels. They are re-scaled logarithmically in the same way (see colourbar at the top). The black cross is the brightest pixel of the hard X-ray image. North is up, east to the left.}
\end{figure*}

We also tried to do some spatially resolved spectroscopy, to look for variations of the properties of the soft X-ray emitting gas along the distance from the nucleus. Following the analysis performed on the RGS high resolution spectra, we chose two narrow energy bands, which we know are dominated by emission from {O\,\textsc{vii}} (0.5-0.6 keV) and {O\,\textsc{viii}} (0.6-0.7 keV)\footnote{By means of simulations with the observed RGS fluxes of the emission lines and the spectral response of the \textit{Chandra} observation, we estimated that the 0.5-0.6 keV band is contaminated by less than 10\% from {O\,\textsc{viii}} photons, while the 0.6-0.7 keV band by around 25\% from {O\,\textsc{vii}} photons.}. As shown in Fig.~\ref{o7_o8}, there is an hint that the {O\,\textsc{vii}} emission is more extended than the {O\,\textsc{viii}} one, but the quality of the data does not allow us to quantify this difference.

We also extracted two spectra, one only of the inner 1 arcsec region, the other excluding it, while keeping all the remaining X-ray emitting region. While practically all the flux above 2 keV is concentrated in the nuclear region, approximately 15\% of the 0.5-2 keV flux is produced farther than 1 arcsec from the nucleus. This percentage is significantly lower than what is generally found in Seyfert 2s observed by \textit{Chandra} \citep[see e.g.][]{bianchi06}. The best fit parameters for the three phases of the soft X-ray emission in the nuclear region are consistent with those found for the outer region. There is an hint of an higher relative flux between the collisional and the photoionisation gas in the outer region with respect to the nucleus, but it is not statistically significant.

\section{Discussion}

\subsection{Fluorescence from Compton-thick material}

The high energy spectrum of Mrk~573 is typical of a Compton-thick Seyfert 2 galaxy. While the Compton reflection component and the strong iron K$\alpha$ line are ubiquitous signatures of reflection from a Compton-thick material, the presence of fluorescent K$\alpha$ lines from neutral silicon and sulphur are less common in spectra of obscured AGN \citep[but see e.g.][]{Sambruna01b,bianchi05b,til08}. This is due to the fact that the fluorescence yield is a strong function of the atomic number, and is thus much lower for low-Z metals with respect to iron. Using the \citet{basko78} formul\ae, valid in the case of a semi-infinite plane-parallel slab isotropically illuminated, we calculated the expected ratios between the fluxes of the silicon, sulphur and iron K$\alpha$ fluorescent lines. We adopted fluorescence yields of 0.042, 0.077 and 0.304, respectively \citep{km93}, and the \citet{ag89} abundances. For all the lines, we considered both the unscattered and the once-scattered photons, since with the quality of our data we cannot disentangle the two components even in the iron line. The calculated ratios are Si/Fe=$0.010-0.028$, S/Fe=$0.014-0.032$, Si/S=$0.70-0.88$, where the ranges take into account the dependencies on the inclination angle and the incident power law index. These values, once transformed in EWs with respect to the incident continuum, are very similar to the ones calculated by \citet{mfr97}.

The observed ratios are Si/Fe=$0.18\pm0.14$, S/Fe=$0.22\pm0.16$, Si/S=$0.8\pm0.8$. The fluxes of the observed silicon and sulphur fluorescent lines are significantly higher than predicted, although the uncertainties are quite large. It is difficult to recover the observed fluxes with elemental overabundances, which would be very high, and apparently not required by the ionised plasma dominating the soft X-ray emission. It is likely that the observed fluxes of these two lines are rather unreliable, because of the low statistics and resolution of the spectra. Indeed, the neutral Si K$\alpha$ is probably contaminated by the {Mg\,\textsc{xii}} K$\beta$ line, at 1.745 keV, which could be underpredicted by our photoionisation model. We note that the fluxes of neutral Si and S K$\alpha$ are also large in other sources where they are tentatively detected \citep[e.g.][]{Sambruna01b,bianchi05b,til08}.

\subsection{The soft X-ray emitting region}

The soft X-ray emission of Mrk~573 appears dominated by a photoionised gas, morphologically coincident with the optical NLR, as commonly found in Seyfert 2 galaxies \citep[e.g.][]{bianchi06,gb07}. From the \textit{Chandra} images, we can see that the two, symmetric, emitting regions have an inner radius $r_i<90 pc$ (pixel size of 0.246 arcsec), and an outer radius $r_o\simeq1$ kpc. The projected opening angle of the cones is roughly 50 degrees, in agreement with the 45 degrees measured by \citet{wt94}. Since the intrinsic luminosity of this source can only be indirectly estimated, because the nucleus is obscured by a Compton-thick material, it is difficult to derive a better guess for the inner radius of the emitting region, from the ionisation parameters of the fits with photoionisation models.

On the other hand, from the luminosity of the strongest line, the {O\,\textsc{vii}} forbidden line (which is not significantly contaminated by the collisional phase), we can calculate its emission measure (EM), and estimate the density of the material producing it. The luminosity (ph s$^{-1}$) of a recombination line of charge state $i$ is:

\begin{equation}
 L_i =\int_V \! A_Z f_{i+1} \eta \alpha(T) n_e n_H \, dV
\end{equation}  

\noindent where $A_Z$ is the abundance of the element, $f_{i+1}$ is the charge state fraction of the recombining ion, $\alpha(T)$ is the radiative recombination coefficient for the recombining ion at temperature $T$, $\eta$ is the fraction of recombinations leading to the relevant transition, and $n_e$ and $n_H$ are the electron and hydrogen density in the volume V of the emitting region \citep[e.g.][]{lied99}. The EM is defined as:

\begin{equation}
 EM\equiv\int_V \! n_e n_H \, dV = \frac{L_i }{A_Z f_{i+1} \eta \alpha(T)}
\end{equation} 

\noindent where the last equivalence assumes that all the relevant parameters are constant in the gas (or their average values are used). Adopting the best-fit parameters of the RGS fit, \textsc{cloudy} gives the following values: $A_Z=4.9\times10^{-4}$, $f_{i+1}=0.13$, $\alpha(T)=1.3\times10^{-11}$ cm$^3$ s$^{-1}$ (for an average temperature of $4.7\times10^4$ K), and $\eta=0.48$. From the observed luminosity of the {O\,\textsc{vii}} forbidden line, we get an EM of $9\times10^{64}$ cm$^{-3}$. The volume of the emitting region may be approximated ($r_i \ll r_o$)  to $V\simeq g [1-\cos(\pi/4)] \pi r_o^3/6 $, where $g$ is the filling factor of the emitting gas within the spherical bi-cone with opening angle of 45 degrees. By assuming $n_e\simeq1.2 n_H$, we derive:

\begin{equation}
 g n_e^2 \simeq 25 \, \mathrm{cm^{-6}}
\end{equation} 

Therefore, the filling factor $g$ is likely to be much smaller than 1. If we take the average densities measured in the optical NLR \citep[of the order of $10^2-10^3$ cm$^{-3}$:][]{cap96,ferr99,schl09}, we derive $g\simeq10^{-3}-10^{-5}$, i.e. the emitting gas fills a very small fraction of the whole volume of the bi-conical region. Such a small filling factor is also required in order to recover a column density of the order of  $3\times10^{20}-6\times10^{21}$ cm$^{-2}$, measured from the RGS and ACIS spectra, along the observed $\simeq1$ kpc. This result is in agreement with the very low mean density ($\simeq1$ cm$^{-3}$), derived from the EM of H$\beta$, if a unity filling factor is assumed \citep{schl09}.

\subsection{The collisional phase and the radio emission}

The detailed spectral analysis of the RGS spectra of Mrk~573 pinpointed the presence of a gas phase in collisional equilibrium, which contributes by $\simeq20-30\%$ to the overall soft X-ray luminosity. The observed thermal emission could be in the hot gas surrounding a starburst region. If all the 0.5-2 keV luminosity emitted by the collisional phase is associated to star formation, we can estimate a star forming rate (SFR) of $\simeq9$ M$_\odot$ yr$^{-1}$ \citep{ranalli03}. A SFR of the same order ($\simeq5$ M$_\odot$ yr$^{-1}$) can be estimated from the total radio luminosity of the two spots, if related completely to the star formation, assuming the relation found by \citet{bell03}, and an average spectral index of $\alpha=-0.75$ \citep{fal98}. A lower SFR ($\simeq2$ M$_\odot$ yr$^{-1}$) is derived from the far infrared (FIR) emission \citep[where the starburst emission largely dominates over the AGN, e.g.][]{fritz06} adopting the relation found by \citet{ken98}, and the FIR luminosity calculated from the \textit{IRAS} infrared fluxes \citep{mosh90} and the method proposed by \citet{hel88}. Moreover, as already reported in Sect.~\ref{hst}, there is no clear evidence of star-forming regions in the inner few kpc in the \textit{HST} near-IR images.

On the other hand, the collisional phase may be directly connected to the observed radio emission, both arising as free-free emission of a hot gas. The X-rays to radio ratio for thermal bremsstrahlung emission of a plasma at temperature $T$ is:

\begin{equation}
 \frac {g\left( \nu_x, T \right) e^{-\frac{h\nu_x}{kT}}} {g\left( \nu_r, T \right) e^{-\frac{h\nu_r}{kT}}}
\end{equation} 

\noindent where $g\left( \nu, T \right)$ are the Gaunt factors appropriate for the X-rays ($\nu_x$) and radio ($\nu_r$) frequency, at that temperature \citep[e.g.][]{longair92}. Adopting a temperature of $5\times10^6$ K, of the order of the one measured for the collisional plasma in the RGS and ACIS spectra, the Gaunt factor at 5 GHz is around 11, while at 1 keV is approximately 1.5 \citep{longair92}. Therefore, the luminosity ratio between the X-rays and the radio free-free emission should be $\simeq0.01$. For a radio luminosity of $10^{28}$ erg s$^{-1}$ Hz$^{-1}$ (representative of each of the regions detected in the VLA data, see Table \ref{radiotable}), we would expect an X-ray luminosity of $10^{26}$ erg s$^{-1}$ Hz$^{-1}$ at 1 keV, i.e. $10^{43.4}$ erg s$^{-1}$ keV$^{-1}$, several orders of magnitude larger than what is observed.

Note that this does not exclude that all the radio emission that we observe in Mrk~573 is due to free-free emission of a hot plasma. The nuclear core, for example, is likely to be absorbed by a substantial column density of neutral gas, like the X-ray nucleus, so that the X-ray emission possibly associated with the radio emission would be completely suppressed. On the other hand, the radio spots, extended on larger scales, should not be affected by such a large obscuration, and cannot be due to thermal bremsstrahlung emission, without overproducing the observed X-ray emission by a large factor. In any case, it is clear that the collisional plasma we detect in the X-rays cannot be produced by a thermal plasma which gives rise also to the radio emission.

The radio ejecta may still be responsible for the heating of the X-ray-emitting plasma, though not emitting in X-rays themselves. In this scenario, the radio luminosity of the two jets is mainly due to non-thermal synchrotron, while the steep power law index \citep{fal98} makes their X-ray luminosity negligible. Indeed, \citet{cap96} assigned a crucial role to the radio ejecta, because they compress the line-emitting gas, enhancing the emission where this interaction occurs. This interaction could also be the heating source of the gas at $\simeq5\times10^6$ K that we observe through its signatures of plasma in collisional equilibrium. This scenario could be tested by making a comparison between the pressure of the hot X-ray emitting gas and the minimum pressure of the radio jets. The latter can be estimated by assuming equipartition between the particles and the magnetic field. We model each jet lobe as a sphere of radius 0.2 kpc. We assume the electron energy distribution extends from a Lorentz factor $\gamma_{\rm min}=2$ up to $\gamma_{\rm max}=10^{5}$, with an electron energy index $p=2.4$ \citep[the results have only a weak dependence on the choice of $\gamma$; see][]{hard04}. Under the simple assumption that the jet is in the plane of the sky, and that relativistic beaming is unimportant, the minimum pressure of each lobe is $\sim 3\times10^{-11}$ barye. On the other hand, the pressure of the hot gas can be estimated from the density directly derived from the \textsc{APEC} normalisation of our \textit{Chandra} best fit. Assuming the same conical emitting volume as in the previous Section, but with an outer radius of 500 pc, similar to the one observed for the radio emission, we get a pressure of $\simeq1.4\times10^{-9}$ barye (a very similar value, $\simeq1.1\times10^{-9}$ barye, is derived if we use the \textsc{APEC} normalisation of our RGS best fit).

The pressure of the hot gas is therefore about 50 times larger than the minimal internal pressure in the radio jets. For the shocks to be the source of heating of the gas, the two pressures should be comparable. In order for the jet not to be suppressed (unless we are not seeing it at some special time), there must be some additional pressure. Indeed, it was suggested that departures from equipartition may characterise the weak jets and lobes observed in Seyfert galaxies and FR I-type radio galaxies \citep[see e.g.][for the case of NGC~2110]{evans06}, differently from the powerful jets observed in FR II-type radio galaxies. However, other studies of the NLR in Mrk~573 have shown that the the optical spectrum is well reproduced by nuclear photoionisation, and there is no evidence of shocks produced by the radio outflows \citep{ferr99,schl09}. The latter do not likely have any strong influence on the NLR medium, apart from some kinematic effects due to their expansion into the gas, eventually influencing the formation of the observed arcs \citep{schl09}.

\section{Conclusions}

We presented a self-consistent analysis of the XMM-\textit{Newton} RGS spectra of the Seyfert 2 galaxy, Mrk~573. Several pieces of evidence suggest that the dominant ionisation process of the soft X-ray emitting gas is photoionisation: the clear detection of RRC from {O\,\textsc{vii}} and {O\,\textsc{viii}}, and the prominence of the {O\,\textsc{vii}} forbidden line. A photoionisation model fully takes into account all the brightest emission lines, but a collisional phase is also required, in order to reproduce the {Fe\,\textsc{xvii}} lines. This component accounts for about 1/3 of the total luminosity in the 15-26 \AA\ band.

The broadband \textit{Chandra} ACIS spectrum confirms the Compton-thick nature of the source, dominated by a Compton reflection component and a strong neutral iron K$\alpha$ line. The soft X-ray emission needs a further photoionisation component with respect to the RGS spectrum, with a larger ionisation parameter, in order to reproduce emission from higher Z metals.

The \textit{Chandra} soft X-ray image closely follow the NLR morphology mapped by the [{O\,\textsc{iii}}] emission. On the other hand, the radio emission is far more compact, although clearly aligned with the NLR. It could be directly related to the collisional phase found in the X-ray spectra, in a plasma heated by the interaction with the radio ejecta, but the estimated pressure of the hot gas is much larger than the pressure of the radio jets, assuming equipartition and under reasonable physical parameters. Alternatively, the gas in collisional equilibrium may originate in a starburst region, requiring a star formation rate of $\simeq5-9$ M$_\odot$ yr$^{-1}$, but there is no clear evidence of this kind of activity from other wavelengths. Deeper X-ray observations are needed in order to confirm the presence of a gas in collisional equilibrium in Mrk~573, and understand its nature.

\section*{Acknowledgements}

SB, EP and GM acknowledge financial support from ASI (grant I/088/06/0). We would like to thank Craig Gordon for support on XSPEC and HEADAS software, Peter Young for support on CHIANTI, Peter van Hoof and Gary Ferland for support on CLOUDY, and Joel H. Kastner for support on the SER procedure. We also thank Alessandro Caccianiga for useful discussions. SB thanks the INAF-OAB for hospitality. CHIANTI is a collaborative project involving the NRL (USA), the Universities of Florence (Italy) and Cambridge (UK), and George Mason University (USA).

\bibliographystyle{mn2e}
\bibliography{sbs}

\label{lastpage}

\end{document}